\documentclass[trackchanges, twocolumn]{aastex701}
\usepackage{comment}
\usepackage{amsmath}

\begin{document}

\linenumbers

\title{Enhanced Detectability of Very-High-Energy Gamma Rays from Off-Axis Short Gamma-Ray Bursts: Impacts of Structured Jets and External Inverse-Compton Scattering}

\author[orcid=0000-0003-2477-9146]{Yuri Sato}
\affiliation{Astronomical Institute, Graduate School of Science, Tohoku University, Sendai 980-8578, Japan}
\email[show]{yuri@astr.tohoku.ac.jp}  

\author[orcid=0000-0003-2478-333X]{Bing Theodore Zhang}
\affiliation{Key Laboratory of Particle Astrophysics and Experimental Physics Division and Computing Center, Institute of High Energy Physics, Chinese Academy of Sciences, Beijing 100049, China}
\affiliation{TIANFU Cosmic Ray Research Center, Chengdu, Sichuan, China}
\email{}

\author[orcid=0000-0002-5358-5642]{Kohta Murase}
\affiliation{Department of Physics, Graduate School of Science, The University of Tokyo, Bunkyo, Tokyo 133-0033, Japan}
\affiliation{Department of Physics; Department of Astronomy \& Astrophysics; Institute for Gravitation and the Cosmos, The Pennsylvania State University, University Park, PA 16802, USA}
\affiliation{Center for Gravitational Physics and Quantum Information, Yukawa Institute for Theoretical Physics, Kyoto University, Kyoto, Kyoto 606-8502, Japan}
\email{}

\begin{abstract}
The binary neutron star merger GW170817 established a direct link between gravitational waves and short gamma-ray bursts (sGRBs), whose afterglows are commonly interpreted as nonthermal emission from structured relativistic jets viewed off-axis. 
Some sGRBs exhibit extended or plateau X-ray emission, suggesting prolonged central-engine activity. 
These X-ray photons can serve as seed photons for external inverse-Compton (EIC) scattering by relativistic electrons in the jet, producing very-high-energy (VHE) gamma rays. 
However, how the angular structure of the jet affects the detectability of EIC emission remains insufficiently explored. 
Here, we investigate EIC emission from off-axis structured jets and assess its detectability with the Cherenkov Telescope Array Observatory (CTAO). 
We find that EIC emission powered by extended and/or plateau X-rays produces detectable VHE signals from nearby sGRBs, particularly during the early afterglow, significantly outshining synchrotron self-Compton emission. 
In our adopted models, Gaussian and power-law jets produce brighter off-axis EIC emission than a top-hat jet, extending the viewing-angle range over which VHE emission may be detectable. 
These results demonstrate that early VHE observations with CTAO can probe both prolonged central-engine activity and the angular structure of relativistic jets in sGRBs.
\end{abstract}

\keywords{\uat{Gamma-ray bursts}{629} --- \uat{Gravitational Waves}{678} ---  \uat{Radiative processes}{2055}}

\section{Introduction} 

Short gamma-ray bursts (sGRBs) are widely believed to originate from the merger of binary neutron stars (BNSs) or a neutron star and a black hole \citep[e.g.,][]{Eichler1989,Narayan1992}. 
The first joint detection of gravitational waves (GWs) and electromagnetic radiation from GW170817/GRB~170817A by the Laser Interferometer Gravitational-Wave Observatory (LIGO) and the Virgo Observatory established a direct connection between sGRBs and compact binary mergers \citep{Abbott2017a,Abbott2017b}. 
Despite this breakthrough, some physical processes including details of nonthermal emission properties of sGRBs remain uncertain.

The X-ray afterglows of sGRBs often exhibit long-lasting emission components.
Many events show extended emission lasting for $\sim10^{2-3}~{\rm s}$ with luminosities of $\sim10^{47-49}~{\rm erg~s^{-1}}$ \citep{Norris2006,Sakamoto2011,Gompertz2013,Berger2014,Kagawa2015,Kisaka2017}, while some also display a plateau phase lasting for $\sim10^{4-5}~{\rm s}$ with luminosities of $\sim10^{44-47}~{\rm erg~s^{-1}}$ \citep{Gompertz2013,Gompertz2014,Kisaka2015,Kisaka2017}.
Although the extended emission has been suggested to originate from prolonged central engine activity, potentially powered by a long-lived magnetar \citep{Dai2006,Metzger2008,Bucciantini2012,Gompertz2013,Rowlinson2013,Gompertz2014,Gompertz2015,Lu2015,Rezzolla2015,Gibson2017} or a black hole \citep{Ioka2005,Barkov2011,Nakamura2014,Kisaka2015}, the physical origin of the plateau emission remains uncertain. One possible scenario is that the outflow producing the extended emission catches up with the decelerating jet responsible for the prompt emission and injects energy into the external shock, thereby producing the plateau emission \citep{Matsumoto2020}.
These long-lasting X-ray components may also serve as seed photons for high-energy emission. \citet{Murase2018} showed that X-ray photons from the extended and/or plateau emission can be upscattered by relativistic electrons accelerated at the external forward shock through the external inverse-Compton (EIC) process, producing very-high-energy (VHE) gamma rays.
Cocoon or kilonova photons may also contribute to the VHE gamma-ray emission through a similar mechanism \citep{Kimura2019,Mei2022}.
Indeed, VHE gamma-ray emission has been detected from several long GRBs \citep{Abdalla2019,MAGIC2019a,MAGIC2019b,LHAASO2023,Cao2023,Abe2024}, and a possible VHE gamma-ray detection has also been reported for the short GRB~160821B \citep{Acciari2021}.
Therefore, VHE gamma-ray observations provide a promising probe of the physical origin of the extended and plateau emission.

The broadband afterglow of GRB~170817A is well explained by a structured jet viewed off-axis rather than by a uniform top-hat jet \citep[e.g.,][]{Alexander2018,Margutti2018,Mooley2018a,Mooley2018b,Ioka2019,Lazzati2018,Urrutia2021,Takahashi2021}.
Since both the synchrotron afterglow and the EIC emission are produced at the external shock, their observational properties depend sensitively on the jet angular structure and the viewing angle. 
Nevertheless, previous studies of EIC emission have mainly adopted the top-hat jet model \citep[e.g.,][]{Toma2009,Murase2010,Murase2011,Murase2018,Kimura2019,Zhang2021}. 
How jet angular structure affects the EIC light curves and the prospects for detecting this emission from off-axis sGRBs therefore remains insufficiently explored.

In this Letter, we investigate how jet angular structure and viewing angle affect the EIC emission powered by extended and plateau X-ray emission, and assess its detectability with the Medium-Sized Telescope of the Cherenkov Telescope Array Observatory (CTAO/MST), which is sensitive around 1~TeV. 
Our calculations are performed using the GRB module of the Astrophysical Multimessenger Emission Simulator ({\sc AMES}; \citet{Zhang2021a,Sato2024,Zhang2025}; see also \citet{Murase2011}\footnote{The public version of {\sc AMES} is available at \url{https://github.com/pegasuskmurase/AMES-GRBAfterglow}.}).
We also estimate the joint GW and VHE gamma-ray detection rates for BNS mergers, connecting our predictions to future multimessenger observations. 
Although VHE gamma rays are attenuated by the extragalactic background light (EBL), sGRBs associated with GW events are expected to occur at relatively low redshifts ($z\lesssim0.04$) \citep{LVK2023,Bom2024}, where EBL attenuation is expected to be modest \citep[e.g.,][]{Murase2007,Rudolph2022}.
Nearby GW associated sGRBs therefore provide promising targets for searches for EIC emission with CTAO/MST.

\section{Model Description}
\label{sec:model}

\begin{figure*}
\centering
\begin{minipage}{0.45\linewidth}
\includegraphics[width=1.0\textwidth]{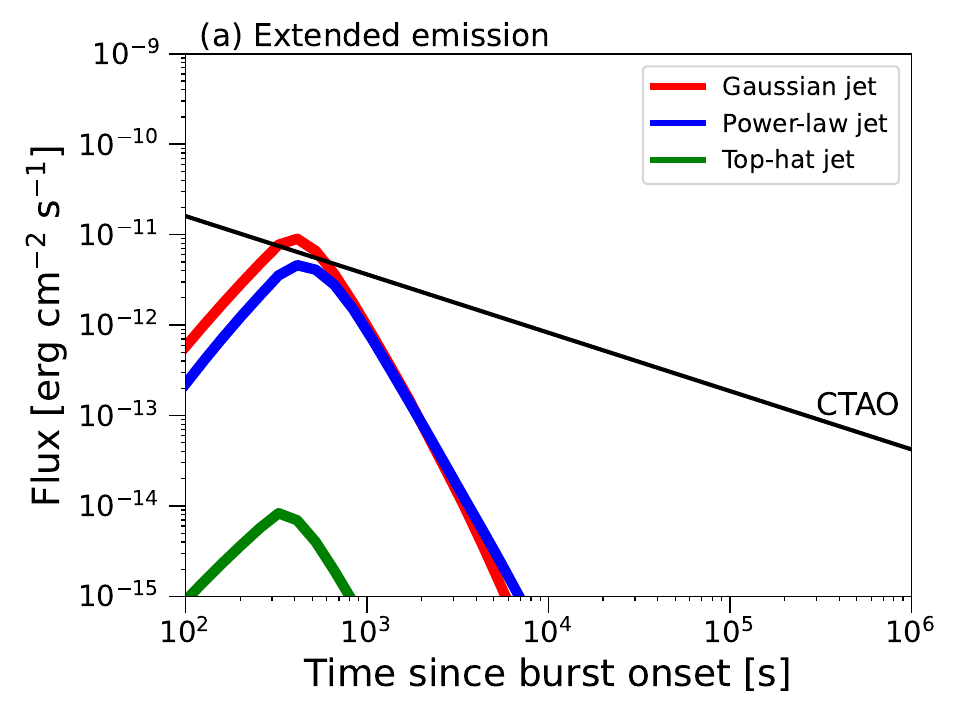}
\end{minipage}
\begin{minipage}{0.45\linewidth}
\includegraphics[width=1.0\textwidth]{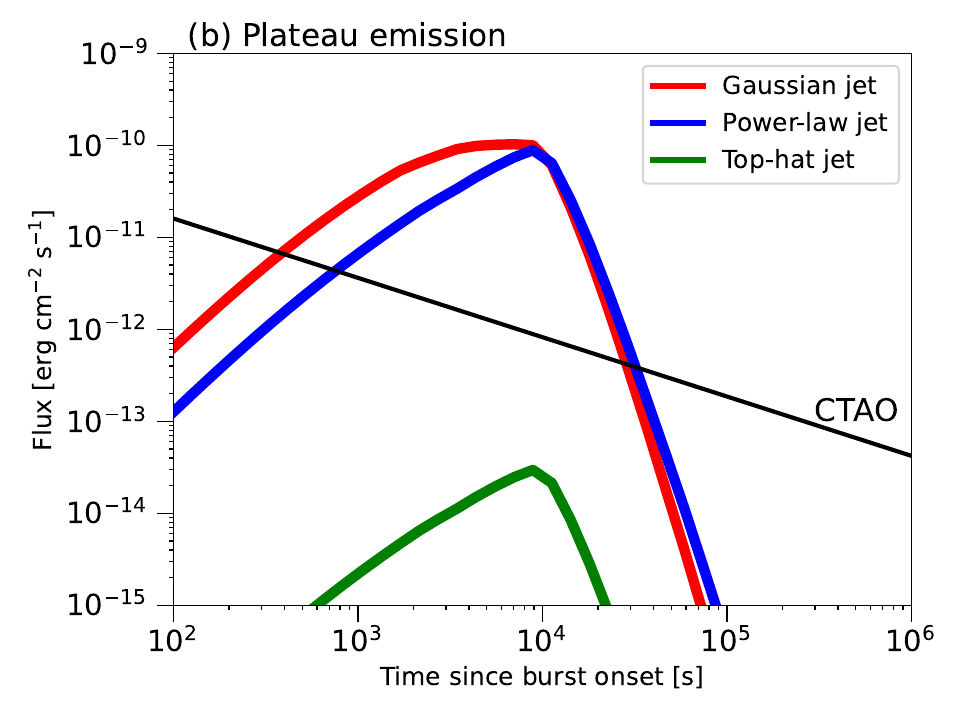}
\end{minipage}
\vspace{-0.05cm}
\caption{
VHE gamma-ray light curves of the EIC emission at $E=1$~TeV for a Gaussian jet (red), a power-law jet (blue), and a top-hat jet (green), viewed at $\theta_v=6^\circ$, assuming a source at $z=0.01$, corresponding to a luminosity distance of $\sim43$~Mpc.
Panel (a) shows the extended emission case, while panel (b) shows the plateau emission case.
The black solid line indicates the CTAO/MST sensitivity \citep{CTA2021}.
}
\label{fig:flux_comp}
\end{figure*}

The angular structure is characterized by the isotropic-equivalent kinetic energy, $\mathcal{E}(\theta)$, and the bulk Lorentz factor, $\Gamma(\theta)$, where $\theta$ is the angle from the jet axis.
The jet opening half-angle is denoted by $\theta_j$.
For the Gaussian jet, we adopt
\begin{equation}
\begin{aligned}
\Gamma_{\rm G}(\theta)&=1+(\Gamma_{c,\rm G}-1)\exp\left(-\frac{\theta^2}{2\theta_{c,\rm G}^2}\right),\\
\mathcal{E}_{\rm G}(\theta)&=\mathcal{E}_{c,\rm G}\exp\left(-\frac{\theta^2}{2\theta_{c,\rm G}^2}\right),
\end{aligned}
\qquad (\theta<\theta_{j,\rm G})
\end{equation}
with $\Gamma_{c, \rm G}=100$, $\mathcal{E}_{c, \rm G}=6\times10^{52}~{\rm erg}$, $\theta_{c, \rm G}=3^\circ$, and $\theta_{j, \rm G}=35^\circ$.
For the power-law jet, we assume
\begin{equation}
\begin{aligned}
\Gamma_{\rm P}(\theta)&=1+\frac{\Gamma_{c,\rm P}-1}{1+\left(\frac{\theta}{\theta_{c,\rm P}}\right)^s},\\
\mathcal{E}_{\rm P}(\theta)&=\frac{\mathcal{E}_{c,\rm P}}{1+\left(\frac{\theta}{\theta_{c,\rm P}}\right)^s},
\end{aligned}
\qquad (\theta<\theta_{j,\rm P})
\end{equation}
with $\Gamma_{c, \rm P}=100$, $\mathcal{E}_{c, \rm P}=6\times10^{52}~{\rm erg}$, $\theta_{c, \rm P}=4^\circ$, $\theta_{j, \rm P}=14^\circ$, and $s=4.7$. 
These parameters are motivated by broadband modeling of GRB~170817A \citep[e.g.,][]{Takahashi2020,Pellouin:2024gqj}. 
For comparison, we also consider an idealized top-hat jet characterized by an initial Lorentz factor of $\Gamma_{\rm t}=164$, an isotropic-equivalent energy of $\mathcal{E}_{\rm t}=6.3\times10^{52}~{\rm erg}$, and a jet opening half-angle of $\theta_{\rm t}=3.4^\circ$.

Each jet propagates into a uniform interstellar medium (ISM) with number density $n_0$. 
The interaction between the relativistic outflow and the ambient medium drives a forward shock, producing synchrotron afterglow emission. 
The shocked electrons are assumed to follow a power-law energy distribution with index $p$. 
We adopt constant microphysical parameters $\epsilon_e$, $\epsilon_B$, and $f_e$, representing the fractions of the post-shock internal energy carried by nonthermal electrons and magnetic fields, and the fraction of accelerated electrons, respectively. 

For VHE gamma-ray radiation, we calculate both synchrotron self-Compton (SSC) and EIC emission using the GRB afterglow module of {\sc AMES} \citep{Zhang2021a,Sato2024,Zhang2025}.
For the EIC component, seed photons from the extended and plateau emission are inverse-Compton scattered by the relativistic electrons accelerated at the forward shock. 
The comoving-frame seed photon spectral energy distribution is described by a broken power law,
\begin{equation}
G(\varepsilon) =
\begin{cases}
\left( \dfrac{\varepsilon}{\varepsilon_b} \right)^{-\beta_1 + 1} & (\varepsilon < \varepsilon_b), \\
\left( \dfrac{\varepsilon}{\varepsilon_b} \right)^{-\beta_2 + 1} & (\varepsilon \geq \varepsilon_b),
\end{cases} 
\label{eq:seed_SED}
\end{equation}
where $\varepsilon$ is the seed photon energy in the comoving frame and $\varepsilon_b=E_b/\Gamma$ is the corresponding break energy.
Following \citet{Murase2018}, we adopt $\beta_1=0.5$ and $\beta_2=2.0$.
The seed photon luminosity evolves as
\begin{equation}
E F_E^{\rm seed}(T) \propto L(T) \propto
\begin{cases}
T^{-\alpha_1} & (T < T_a), \\
T^{-\alpha_2} & (T \geq T_a),
\end{cases}
\label{eq:seed_L}
\end{equation}
with $\alpha_1=0$ and $\alpha_2=40/7$ \citep{Kisaka2017,Murase2018}. 
For the extended emission, we adopt $E_b=1\,{\rm keV}$, $L_{E_b}^{\rm seed}=10^{49}\,{\rm erg\,s^{-1}}$, and $T_a=300\,{\rm s}$, while for the plateau emission we use $E_b=0.1\,{\rm keV}$, $L_{E_b}^{\rm seed}=10^{47}\,{\rm erg~s^{-1}}$, and $T_a=1\times10^{4}\,{\rm s}$, following \citet{Murase2018}.

BNS mergers are expected to produce thermal kilonova emission, as observed in GW170817 \citep{Tanvir:2017pws}.
In addition to the extended and plateau emissions, we evaluate the possible contribution of kilonova photons as an external seed photon field at late times.
The possibility of inverse-Compton emission from the interaction of a late time jet with kilonova photons has also been discussed for GRB 211211A \citep{Mei2022}.
We adopt a blackbody spectrum with a fixed temperature of $T_{\rm KN}=5000~{\rm K}$ and a fixed photospheric radius of $R_{\rm KN}=10^{15}~{\rm cm}$, based on observational estimates for GW170817 at around one day after the NS merger \citep{Waxman:2017sqv}. 
Unlike the extended and plateau emission components, whose temporal evolution is considered in our model, we treat the kilonova photon field as time-independent because its temperature and photospheric radius at other epochs are not sufficiently constrained.
These parameters yield a bolometric luminosity of
$L_{\rm KN}=4\pi\sigma_{\rm SB}R_{\rm KN}^2T_{\rm KN}^4$,
where $\sigma_{\rm SB}$ is the Stefan--Boltzmann constant.

The synchrotron, SSC, and EIC emission are calculated by integrating the emissivity over the equal-arrival-time surface. 
Klein-Nishina suppression is fully taken into account following \citet{Nakar2009,Murase2010,Murase2011,Wang2010,Jacovich:2020teu,Sato2023a,Zhang2021a}. 

\begin{figure*}
\centering
\begin{minipage}{0.45\linewidth}
\includegraphics[width=1.0\textwidth]{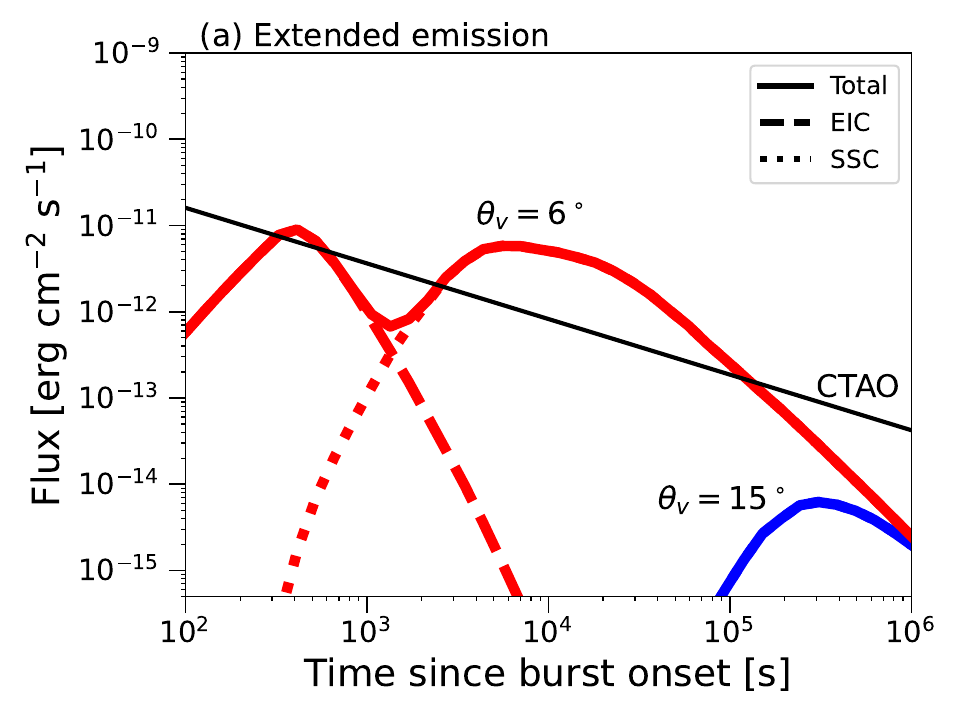}
\end{minipage}
\begin{minipage}{0.45\linewidth}
\includegraphics[width=1.0\textwidth]{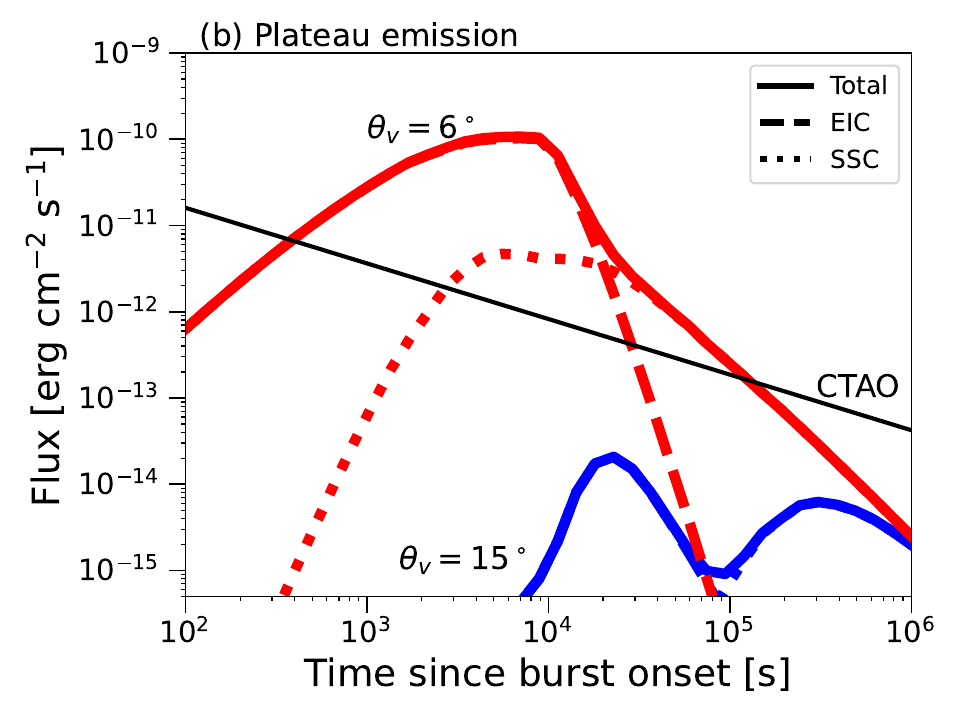}
\end{minipage}
\vspace{-0.05cm}
\caption{
VHE gamma-ray light curves at $E=1$~{TeV} for a Gaussian jet viewed at $\theta_v=6^\circ$ (red) and $15^\circ$ (blue), assuming a source at $z=0.01$, corresponding to a luminosity distance of $\sim43$~Mpc. 
Panel (a) shows the extended emission case, while panel (b) shows the plateau emission case. 
The solid, dashed, and dotted curves represent the total, EIC, and SSC components, respectively. 
The black solid line indicates the CTAO/MST sensitivity \citep{CTA2021}.
}
\label{fig:flux_g}
\end{figure*}
\begin{figure*}
\begin{minipage}{0.43\linewidth}
\centering
\includegraphics[width=\textwidth]{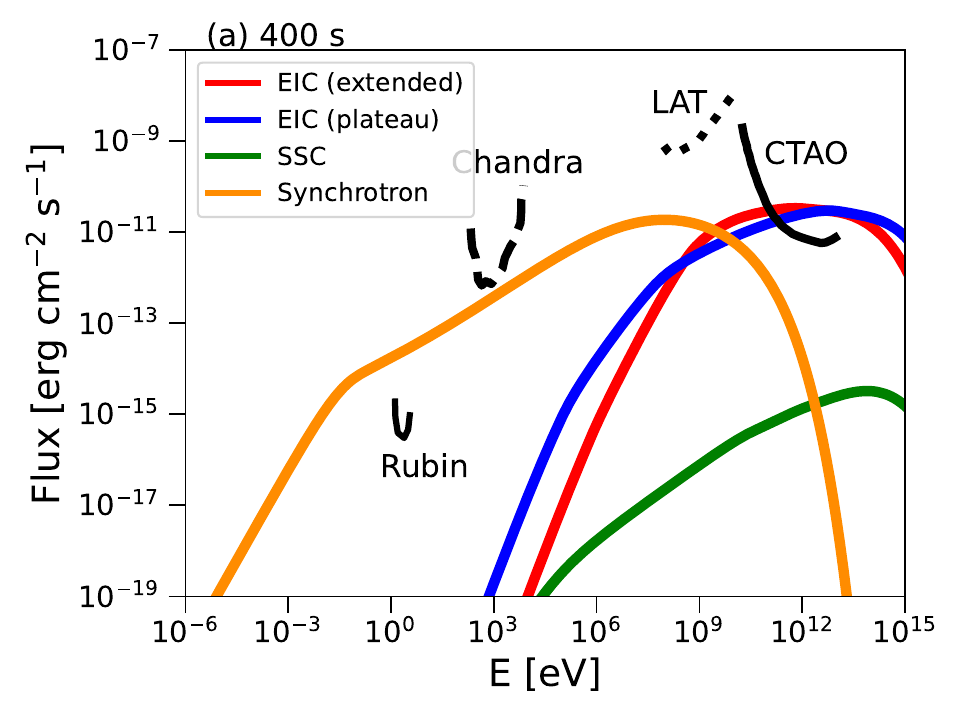}
\end{minipage}
\begin{minipage}{0.43\linewidth}
\centering
\includegraphics[width=\textwidth]{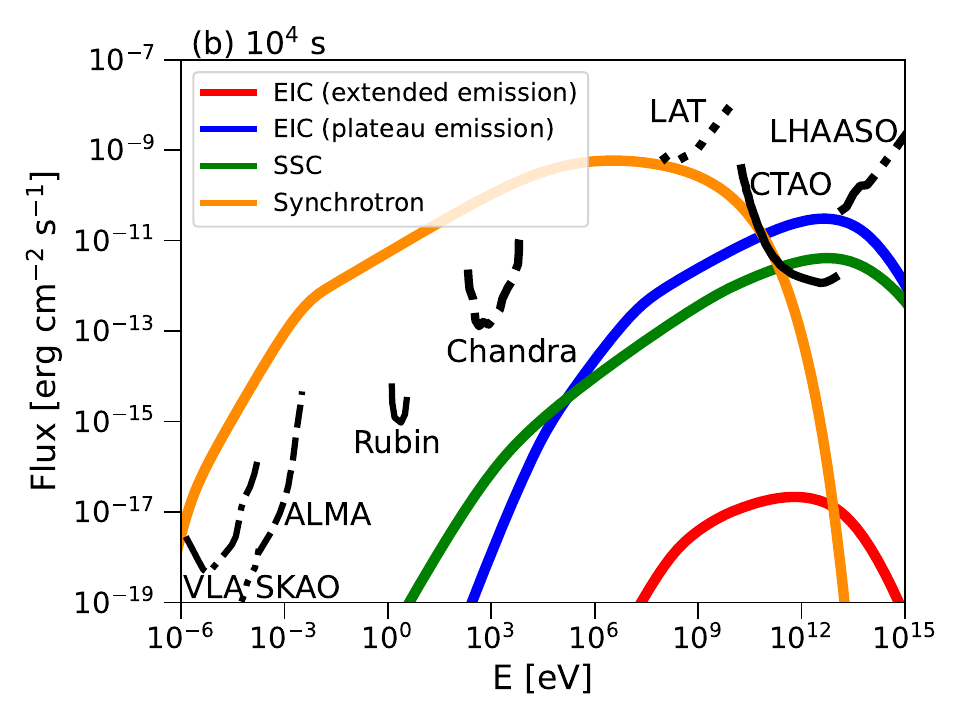}
\end{minipage}
\makebox[\textwidth][c]{%
\begin{minipage}{0.43\linewidth}
\centering
\includegraphics[width=\textwidth]{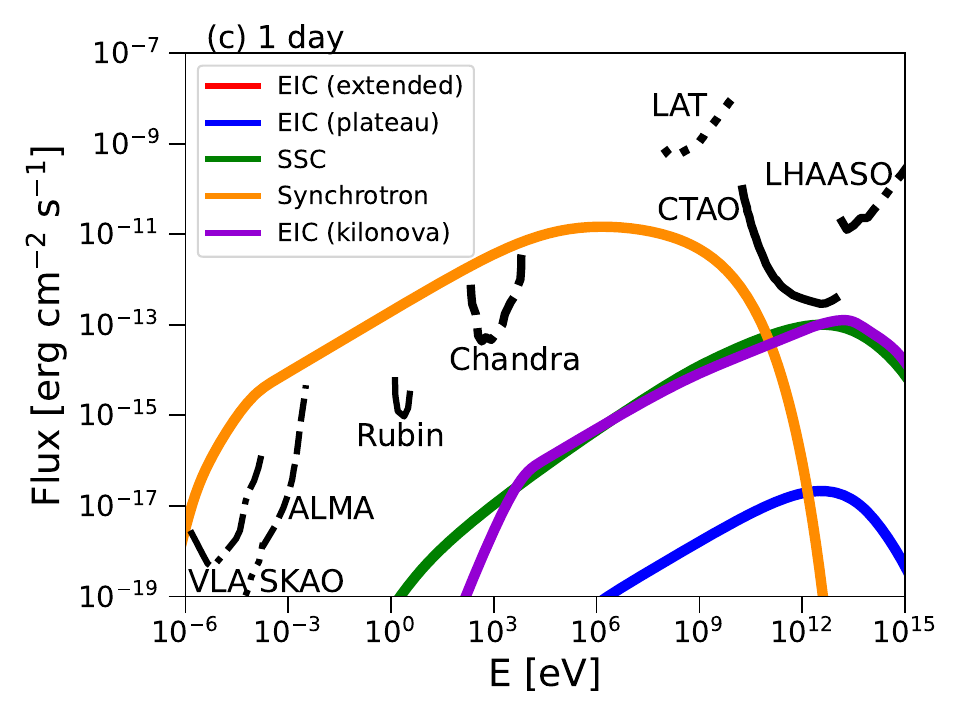}
\end{minipage}
}
\vspace{-0.05cm}
\caption{
Model spectra from a Gaussian jet viewed off-axis at $\theta_v=6^\circ$, shown at $T=400~\mathrm{s}$ (panel (a)), $1\times10^4~{\rm s}$ (panel (b)), and $1~\mathrm{day}$ (panel (c)). The red and blue curves represent the EIC emission associated with the extended and plateau X-ray emission, respectively; the green and orange curves show the SSC and synchrotron components, respectively. The purple curve in panel (c) represents the EIC emission produced by inverse-Compton scattering of kilonova photons.
The black curves indicate the sensitivity limits of CTAO/MST (thick solid; exposure time matched to the epoch shown in each panel; \citet{CTA2021}), LHAASO (thick dot-dashed; $10^4~\mathrm{s}$ in panel (b) and $1~\mathrm{day}$ in panel (c); \citet{LHAASO2025}), \textit{Fermi}/LAT (thick dotted; $100~\mathrm{s}$; \url{https://fermi.gsfc.nasa.gov}), \textit{Chandra} (thin dashed; exposure time matched to the epoch shown in each panel; \url{https://chandra.harvard.edu}), Rubin Observatory (thin solid; $30~\mathrm{s}$; \url{https://www.lsst.org}), ALMA (thin dotted; $1~\mathrm{hr}$; \url{https://alma-telescope.jp/en/}), SKAO (thin dashed; $1~\mathrm{hr}$; \url{https://www.skao.int/en}), and VLA (thin dot-dashed; $1~\mathrm{hr}$; \url{https://public.nrao.edu/telescopes/vla/}). 
The radio facilities (ALMA, SKAO, and VLA) and LHAASO are not shown in panel (a).
}
\label{fig:SED}
\end{figure*}
\begin{figure}
\centering
\includegraphics[width=0.4\textwidth]{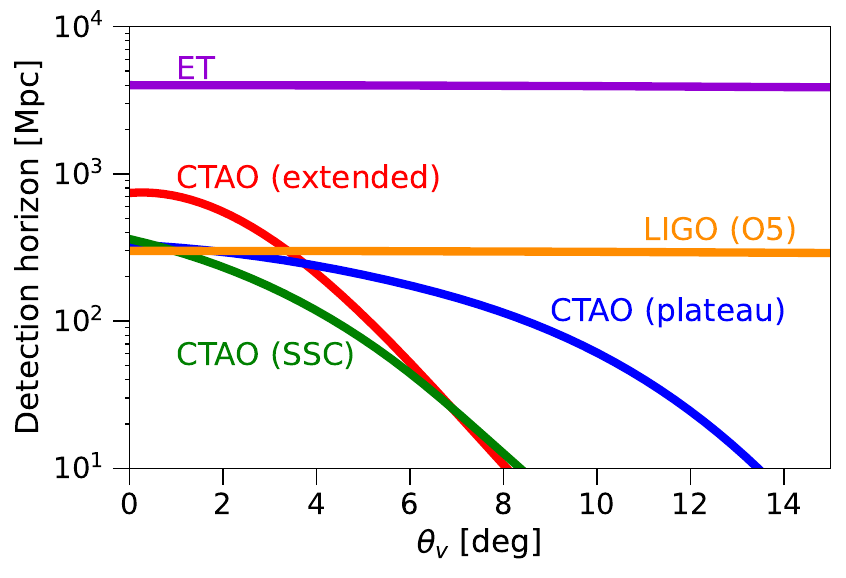}
\vspace{-0.05cm}
\caption{
The GW and CTAO/MST detection horizons are shown as functions of the viewing angle $\theta_v$. 
For Advanced LIGO O5, we adopt a sky averaged BNS detection range of $\bar{H}_{\rm BNS}=300\,\mathrm{Mpc}$ \citep{Mondal:2024irw}, while for the ET, we adopt a representative sky averaged BNS detection range of $\bar{H}_{\rm BNS}=4000\,\mathrm{Mpc}$ \citep{ET2026}. 
The CTAO/MST detection horizon is estimated from the peak flux of the Gaussian jet emission. 
The red and blue curves correspond to the EIC emission in the extended emission and plateau emission cases, respectively, while the green curve represents the SSC emission. 
The orange and violet curves indicate the BNS detection ranges of Advanced LIGO O5 and the ET, respectively.
}
\label{fig:horizon}
\end{figure}

\section{Results of VHE gamma-ray emission}
\label{sec:result}

\subsection{Comparison of Jet Structures}
\label{sec:result_comp}

We first investigate how the jet angular structure affects the VHE gamma-ray emission. 
Unless otherwise noted, we adopt the afterglow parameters $\theta_v=6^\circ$, $\epsilon_e=4.0\times10^{-2}$, $\epsilon_B=2.1\times10^{-5}$, $f_e=1.0$, $p=2.17$, and $n_0=6.3\times10^{-3}~{\rm cm^{-3}}$ motivated by the broadband modeling of GRB~170817A by \citet{Takahashi2020}. 
We assume a redshift of $z=0.01$, corresponding to a luminosity distance of $\sim43$~Mpc, which is close to the peak of the joint GW--CTAO detection probability for BNS mergers \citep{Mondal:2024irw}. 
Throughout this work, we compare the predicted VHE gamma-ray emission with the CTAO/MST sensitivity.

Figure~\ref{fig:flux_comp} shows the 1~TeV EIC light curves for the Gaussian, power-law, and top-hat jets.
The Gaussian and power-law jets produce significantly brighter EIC emission than the top-hat jet for both the extended emission and plateau emission cases.
In contrast, the EIC emission from the top-hat jet is strongly suppressed for off-axis observers because of relativistic beaming effect.
The Gaussian and power-law jets exhibit similar EIC light curves, since their isotropic-equivalent energies and Lorentz factors at $\theta\simeq\theta_v=6^\circ$ are comparable.
Given its slightly higher EIC flux in our adopted parameter setup, we adopt the Gaussian jet as the fiducial model for the remainder of this Letter.

\subsection{Temporal Evolution of EIC and SSC Emission}
\label{sec:result_detect}

Having established the Gaussian jet as our fiducial model, we examine the temporal evolution of the EIC and SSC emission and assess how their detectability depends on the viewing angle.
Unless otherwise noted, the model parameters are identical to those adopted in Section~\ref{sec:result_comp}.
We consider two representative viewing angles, $\theta_v=6^\circ$ and $15^\circ$.
Figure~\ref{fig:flux_g} shows the 1~TeV light curves for these viewing angles together with the CTAO/MST sensitivity.
The total VHE gamma-ray emission is decomposed into SSC and EIC components.

In the extended emission case, the EIC component reaches its maximum at $T\sim300$~s, corresponding to the end of the extended emission phase, and dominates the VHE gamma-ray emission during the early afterglow.
For $\theta_v=6^\circ$, the peak EIC flux exceeds the CTAO/MST sensitivity, whereas for $\theta_v=15^\circ$ it remains below the detection threshold.
The plateau emission case exhibits similar behavior, except that the EIC peak is delayed to $T\sim10^{4}$~s, reflecting the longer duration of the seed photon injection.
For $\theta_v=6^\circ$, the peak EIC flux exceeds the CTAO/MST sensitivity, whereas for $\theta_v=15^\circ$ it remains below the detection threshold, as in the extended emission case.
The SSC component also exceeds the CTAO/MST sensitivity for $\theta_v=6^\circ$.

Distinguishing between the EIC and SSC emission components requires observations during the early afterglow phase, as the two components exhibit different temporal evolution during their rising and declining phases. 
In the structured jet scenario, an off-axis observer effectively views the emitting region at $\theta \simeq \theta_v$ as an on-axis emitter. 
For the viewing angle of $\theta_v=6^\circ$, this region remains in the coasting phase until $T\sim5\times10^3~{\rm s}$.
During this phase, we consider the spectral regime $\varepsilon_m^{\rm EIC}<\varepsilon<\varepsilon_c^{\rm EIC}$, where $\varepsilon_m^{\rm EIC}\sim\gamma_m^2\varepsilon_b$ and $\varepsilon_c^{\rm EIC}\sim\gamma_c^2\varepsilon_b$, with $\gamma_m$ and $\gamma_c$ denoting the minimum and cooling electron Lorentz factors, respectively. 
For $T<T_a$, the seed photon luminosity remains approximately constant, $L(T)\propto T^0$.
Combining Equations~(\ref{eq:seed_SED}) and (\ref{eq:seed_L}) then gives $F_\varepsilon^{\rm EIC}\propto T$ in the Thomson limit. 
At $T\gtrsim T_a\sim300~{\rm s}$, the seed photon luminosity begins to decline as $L(T)\propto T^{-40/7}$, causing the EIC flux to decrease rapidly as $F_\varepsilon^{\rm EIC}\propto T^{-33/7}$.
In contrast, in the SSC regime, $\varepsilon_m^{\rm SSC}<\varepsilon<\varepsilon_c^{\rm SSC}$, where $\varepsilon_m^{\rm SSC}\sim\gamma_m^2\varepsilon_m$ and $\varepsilon_c^{\rm SSC}\sim\gamma_c^2\varepsilon_c$, with $\varepsilon_m$ and $\varepsilon_c$ denoting the characteristic and cooling synchrotron photon energies, respectively, the SSC flux increases as $F_\varepsilon^{\rm SSC}\propto T^4$ throughout the coasting phase, up to $T\sim5\times10^3~{\rm s}$ \citep{Fraija2022}. 
The different temporal indices of the EIC and SSC fluxes during their rising phases therefore provide a potential observational diagnostic for distinguishing between the two emission mechanisms.
The two components also peak at different times because their temporal evolution is governed by different physical processes. 
The EIC emission peaks around $T\sim T_a$, when the seed photon luminosity begins to decline, whereas the SSC emission continues to rise throughout the coasting phase of the emitting region at $\theta\simeq\theta_v$. 
The SSC component is therefore expected to peak around the onset of deceleration of this region. 
Since this occurs later than $T_a$ in our model, the SSC component peaks later than the EIC component and eventually becomes dominant.

\subsection{Broadband Spectra}

To assess the detectability of the predicted emission across different wavelength bands, we compare the broadband spectral energy distributions of our fiducial Gaussian jet model, viewed at $\theta_v=6^\circ$, with the sensitivities of CTAO/MST, the Large High Altitude Air Shower Observatory (LHAASO), the {\it Fermi} Large Area Telescope (\textit{Fermi}/LAT), the {\it Chandra} X-ray Observatory ({\it Chandra}), the Vera C. Rubin Observatory (Rubin), the Atacama Large Millimeter/submillimeter Array (ALMA), the Square Kilometre Array Observatory (SKAO), and the Very Large Array (VLA). 
Figure~\ref{fig:SED} shows the predicted spectra at different epochs together with the corresponding instrumental sensitivities.
The synchrotron component dominates from the radio to X-ray bands, whereas the SSC and EIC components become prominent at VHE gamma-ray energies. During the extended and plateau emission phases, the EIC component produces a distinct VHE gamma-ray excess above the SSC emission, while at later times the EIC emission associated with kilonova photons emerges in the VHE gamma-ray band. 
Comparison with representative instrumental sensitivities illustrates the potential for complementary multiwavelength observations across a broad energy range. In particular, CTAO/MST may detect the VHE gamma-ray emission associated with the EIC component in nearby events, while simultaneous X-ray, optical, and radio observations can help characterize the underlying synchrotron afterglow.
Such coordinated observations provide a promising means of probing long lasting central engine activity and identifying the origin of the VHE gamma-ray emission.

\section{GW--VHE gamma-ray Detectability}

The prospects for detecting GRBs with the CTAO/MST have been investigated in previous studies \citep[e.g.][]{Kakuwa2012}.
We assess the prospects for joint GW and VHE gamma-ray detections of BNS mergers by comparing the CTAO/MST detection horizon with the GW detection ranges and estimating the corresponding joint detection rates. 
Following the methodology of \citet{Mondal:2024irw}, we adopt sky-averaged BNS detection ranges of $\bar{H}_{\rm BNS}=300\,{\rm Mpc}$ and $4000\,{\rm Mpc}$ for Advanced LIGO during the O5 observing run and the next-generation Einstein Telescope (ET), respectively \citep{Mondal:2024irw,ET2026}. 
Figure~\ref{fig:horizon} shows the CTAO/MST detection horizon and the GW detection ranges as functions of the viewing angle. The CTAO/MST detection horizon is estimated from the peak EIC flux for the fiducial Gaussian jet model.
For the extended emission case, the EIC detection horizon decreases rapidly beyond the jet core angle, $\theta_{c,\rm G}=3^\circ$, as both the isotropic-equivalent energy and Lorentz factor decline away from the jet axis. 
The plateau emission case remains detectable to somewhat larger viewing angles because its seed photon field persists until $T_a\simeq1\times10^4\,{\rm s}$, partially compensating for the weaker EIC emission at larger viewing angles.
However, at $\theta_{\rm v}\gtrsim8^\circ$, the Lorentz factor of the material along the line of sight decreases to $\Gamma\sim$ a few, while its isotropic-equivalent energy is also substantially reduced, causing the detection horizon to decline rapidly at larger viewing angles. 
Overall, the CTAO/MST detection horizon is more restricted in viewing angle than the GW detection ranges of Advanced LIGO O5 and ET. 
Joint GW--VHE detections are therefore expected to preferentially select sGRBs viewed relatively close to the jet axis.

We further estimate the expected joint GW--VHE detection rate by integrating the effective detection volume over the viewing angle following \citet{Mondal:2024irw}, using the detection horizons shown in Figure~\ref{fig:horizon}. 
Adopting a local BNS merger rate of $R_{\rm BNS}=10$--$1700~{\rm Gpc^{-3}\,yr^{-1}}$ \citep{Mondal:2024irw}, we obtain joint detection rates of $\sim0.003$--$0.5~{\rm yr^{-1}}$ and $\sim0.004$--$0.7~{\rm yr^{-1}}$ for the EIC emission in the extended and plateau emission cases, respectively, during the LIGO O5 run. 
For comparison, we estimate the joint detection rate based on the SSC emission in our fiducial model, obtaining $\sim0.001$–$0.2~{\rm yr^{-1}}$, which is broadly consistent with the SSC based estimate of \citet{Mondal:2024irw}.
For ET, the joint detection rate is determined primarily by the VHE gamma-ray detectability (see Figure~\ref{fig:horizon}). 
We obtain $\sim0.01$--$2~{\rm yr^{-1}}$ and $\sim0.004$--$0.7~{\rm yr^{-1}}$ for the EIC emission in the extended and plateau emission cases, respectively, while the corresponding rate for SSC is $\sim0.001$--$0.2~{\rm yr^{-1}}$. 
Thus, the joint detection rate for the extended emission case can be about an order of magnitude higher than that for SSC in the ET era. 
The increase in the extended emission rate from LIGO O5 to ET reflects the removal of the GW detection range limitation at small viewing angles, where the EIC horizon exceeds the LIGO O5 GW range.
In contrast, the plateau emission rate changes little because its EIC horizon remains below the LIGO O5 GW range over essentially the entire relevant viewing angle range.

Our results highlight the potential of joint GW--VHE observations to probe the central engines of sGRBs. 
While GW observations identify the merger, early time VHE gamma-ray observations with CTAO/MST can reveal EIC emission powered by prolonged X-ray activity, providing a signature of long lived central engine activity. 
Thus, combining GW detections with rapid VHE gamma-ray follow-up can link the merger to the early high-energy afterglow and simultaneously probe the jet structure and central engine activity. 
This motivates rapid CTAO observations of nearby GW detected BNS mergers, particularly within the first $10^3$--$10^4$~s after the merger.

\section{Summary}

We have investigated EIC emission from structured jets in sGRBs with prolonged X-ray emission and assessed its detectability with the CTAO. 
In our models, Gaussian and power-law jets produce brighter off-axis EIC emission than a top-hat jet, owing to their angular distributions of kinetic energy and Lorentz factor. 
For nearby sGRBs viewed at relatively small off-axis angles, the EIC component can dominate over SSC emission during the early afterglow and may be detectable with CTAO/MST. 
The distinct temporal evolution of EIC and SSC emission provides a potential observational diagnostic of the prolonged seed photon supply and, hence, central-engine activity. 
We have further estimated the joint GW and VHE detection rates for binary neutron star mergers. We have shown that the extended-emission-powered EIC component can yield a joint detection rate approximately an order of magnitude higher than that of the SSC component in the ET era. 
The previous work showed the importance of EIC emission compared to SSC emission~\citep{Murase2018}, but we stress that it is important to consider structured jets in the presence of external seed photons for off-axis observations of VHE gamma rays. These results demonstrate that rapid VHE follow-up observations of nearby GW detected mergers can provide a complementary probe of both prolonged central engine activity and the angular structure of sGRB jets.

\section*{Acknowledgements}
We thank 
Shigeo~S.~Kimura,
Shota~Kisaka,
Junko~Kushida,
Asuka~Kuwata,
Riku~Kuze,
Riki~Matsui,
Masaomi~Tanaka,
Kenji~Toma, and
Ryo~Yamazaki
for valuable comments.
This research has made use of the CTA instrument response functions provided by the CTA Consortium and Observatory, see https://www.ctao-observatory.org/science/cta-performance/ (version prod5 v0.1; \citet{CTA2021}) for more details.
This research was partially supported by JSPS KAKENHI, Grant No.~25KJ0010 (Y.S.). 
The work of K.M. was partially supported by the NSF Grant Nos. AST-2108466, AST-2108467 and AST-2308021.

\bibliography{main_eic}{}
\bibliographystyle{aasjournalv7}

\end{document}